# Digital capabilities assessment for supporting the transformation of the customer experience
# (Preprint Version)

Leonardo Muñoz Muñoz[1*], and Oscar Ávila[1]

[1]Department of Systems and Computing Engineering, School of Engineering,

Universidad de los Andes, Bogotá, Colombia

*To whom correspondence should be addressed; E-mail: l.munozm@uniandes.edu.co;
oj.avila@uniandes.edu.co

**Most of organizations are increasingly investing huge amounts of money today in order to have the right digital capabilities required for their industry. The area where organisations feel they have made the most progress is in improving the customer experience, which encompasses aspects such as data analytics, social media, location-based marketing, mobile channels among others. This aspect became the most important for the survival of organisations since the outbreak of the Covid-19 pandemic. While much has been achieved, many organisations are still not satisfied. One of the major problems in moving forward is the lack of literature in both academia and industry on maturity models allowing organisations to understand their current state in terms of digital capabilities to engage with customers, as well as to plan the evolutionary path to improve in this area. To fulfil this lack, this paper presents the design and validation of a maturity model that enables organizations to assess their digital capabilities in order to improve customer experience and engagement throughout the customer lifecycle.**

# Introduction

The use of new technological and digital trends such as the Internet of Things (IoT), hybrid and augmented reality, machine learning, among others, are changing the way people work and go about their daily activities, how governments serve citizens and how companies do business. Organisations have realised that implementing such technology trends can enable them to create new digital and business capabilities to improve their operational performance and meet the needs of external stakeholders (Adam et al., 2020). The positive impact of new technologies is not limited to planned changes, also in non-planned critic situations, as the lock down driven by COVID crisis, the new technologies were the main pilar to guarantee organisations' continuity.

There are several definitions of Digital Transformation (DT). According to (Vial, 2019), it is "a process that aims to improve an entity by triggering significant changes to its properties through combinations of information, computing, communication, and connectivity technologies". According to (Morakanyane et al., 2017), DT is "an evolutionary process that leverages digital capabilities and technologies to enable business models, operational processes and customer experiences to create value". In both definitions, DT is about bringing improvement, evolution and change to the organisation through digital, information and communication technology implementation.

One of the main impacts of the DT initiatives refers to the development of digital capabilities enabled by Information Technologies (IT). In the private sector, an IT-enabled capability refers to an organisational capability implemented by using and leveraging IT to differentiate from competitors (Sambamurthy & Zmud, 2000). An organisational capability refers to the ability to acquire or develop knowledge for coordinating resources, executing processes, routines and activities, and developing the skills that the firm requires. Previous research has found evidence that capabilities enabled by digital technologies lead to better organisational performance (Ayabakan et al., 2017) and is a driver of business value (Benitez et al., 2018).

There are three key business pillars in which organisations are driving the creation of digital capabilities in order to achieve DT (MIT-Capgemini, 2018; Muñoz, 2018; Westerman G. et.al., 2014a, 2014b). These areas are improving customer experience, driving operational excellence, and adjusting or even reinventing their business model. The customer experience is composed of cognitive, emotional, physical, sensory, spiritual and social elements that mark the direct or indirect interaction of the customer with the organisation (De Keyser et al., 2015). In this pillar the generation of digital capabilities includes both getting to know the customer intimately which refers to the use of tools to understand the customer behaviour and needs, and designing the customer experience that relates to delivering omni-channel experience, product personalization, self-service tools, among others. Operational excellence includes improving the performance of industrial and business processes. Performance is mainly articulated through two concepts, effectiveness which consists of producing a desired result or achieving a defined goal, and efficiency that refers to executing an activity or producing a result with the least possible resources. Finally, the transformation of the business model includes deeper changes than those mentioned in the previous pillars as it involves the use of new technological trends to turn the organisation into a disruptive player in its industry. More recently, digital capabilities also encompass a new pillar called talent and organization that reflects the growing need for organizations to involve their employees in DT and adapt their structure to the demands of a digital organization.

From a recent study (MIT-Capgemini, 2018), only 39% of organizations on average say they have the right digital capabilities required for their industry. The area where organisations feel they have made the most progress is in improving the customer experience, which encompasses aspects such as data analytics, social media, location-based marketing, mobile channels and connected products. This aspect is becoming the most important for the survival of organisations since the outbreak of the Covid-19 pandemic because of the need to digitise many of the interactions that were previously done physically with customers. While much has been achieved, many organisations are still not satisfied. One of the major problems in moving forward is the lack of literature in both academia and industry on maturity models allowing organisations to understand their current state in terms of digital capabilities to engage with customers, as well as to plan the evolutionary path to improve in this area. To fulfil this lack, this paper presents the design and validation of a maturity model that enables organizations to assess their digital capabilities in order to improve customer experience and engagement throughout the customer lifecycle.

This article is structured as follows: Chapter *"Foundation"* presents conceptual foundation of our research, chapter *"Research and construction process"* presents our research approach for the construction of the model, chapter *"Hypothesis generation"* presents the generation of hypothesis from the analysis of the research literature, chapters *"Interviews with domain experts"* presents the validation of the hypothesis through interviews with domain experts, chapter *"Refinement of the model"* illustrates the design of the model, chapter *"Validation: practical application"* describes the validation of the model through three case studies, chapter *"Discussion"* describes the discussion of our contribution and chapter *"Conclusions and future work"*.

## Foundation

### Customer experience

The customer experience can be defined as "comprised of the cognitive, emotional, physical, sensorial, spiritual, and social elements that mark the customer's direct or indirect interaction with (an)other market actor(s)" (De Keyser et al., 2015). In general, for a company to improve or manage the customer experience two main activities are necessary: first, understanding the customer needs and behaviour, and second, designing the customer experience through different digital and traditional channels (De Keyser et al., 2015; MIT-Capgemini, 2018; Morakanyane et al., 2017; Westerman G. et.al., 2014a, 2014b). Regarding the first activity, practitioners and academicians agree that understanding the customer experience concerns several dimensions that include emotional, cognitive, behavioural, social and sensorial components that should be evaluated through the contacts between the firm and the customer at different touch points through the customer journey (Homburg et al., 2017). Concerning the second activity, customer experience design includes choosing adequate channels and touchpoints depending on the behaviour of the customer, her preferences and needs to support the interaction of the customer and the firm as well as selecting the characteristics of the technology offered to support such interaction (Mosa et al., 2020). On this basis, the maturity model proposed in this research aims at assessing digital capabilities and capabilities an organization possesses in order to carry out both activities, customer understanding and customer experience design.

### Maturity models

Maturity models can help understand the current state and gradual improvement of general skills, processes, structures, or conditions of an organization (Blondiau et al., 2016). When introducing transformation or changes to an organization, it is important before to assess its current state or the evolution it has had as well as plan possible path or ways to its evolution. This is one reason why maturity models are used when processes, capabilities, structure or others are being improved. In other words, this type of model can be used as a multi-stage planning tool to identify which improvements should be introduced and when from the as-is state of the organization. To do so, first, the process or capability that must be improved is assessed based on the maturity model. Then, the evaluation outcome is used to identify which improvements must be introduced to increase the organization's maturity level (Helgesson et al., 2012).

In this context, the model designed in this paper follows a top down approach in which the maturity levels are first defined and then the assessment criteria corresponding to each level are outlined (Santos-Neto & Costa, 2019). In addition, the model is descriptive, so that its application consists of a single specific evaluation allowing the organization to understand the situation as it is. Lastly, in the proposed model, maturity for each criterion is represented as a series of one-dimensional linear stages. At the end, the final result is an 'average' maturity stage being provided to the assessed entity, which has been widely accepted and used to as an outcome for maturity models (Santos-Neto & Costa, 2019).

## Research and construction process

The research process (see Figure 1) was derived from the process for reference model construction by (Ahlemann, 2009) that comprises the following four phases:

1. Problem definition. The research problem was defined and documented in the introduction section.
2. Hypotheses generation. The second phase consists of two different activities:

   a. Analysis of the research literature: we consider further research in the academic literature on customer experience and relationship to generate hypotheses on the domain. This is presented in section with the same name below.

b. Design of the maturity model structure. The initial construction of the maturity model structure is based on the knowledge obtained from the results of the literature review (step 2.a.). This is described in section *"Towards a model for digital maturity assessment"*.

3. Validation. The objective of this phase is to validate and complete the construction of the model. It is made through three activities:

a. Interviews with domain experts. Four experts in DT were interviewed with the aim of gathering further empirical evidence. Two activities were approached from the discussions held with the interviewees: the refinement of the model and its validation. The results of both activities are described in sections *"Interviews with domain experts"* and *"Refinement of the model"* respectively.

b. Practical application. The validation of the model was also achieved through his application to a concrete case study (see section *"Validation: practical application"*).

4. Documentation. The documentation of the model contains a description of the construction process, the model itself, the documentation of the interview and case study results.

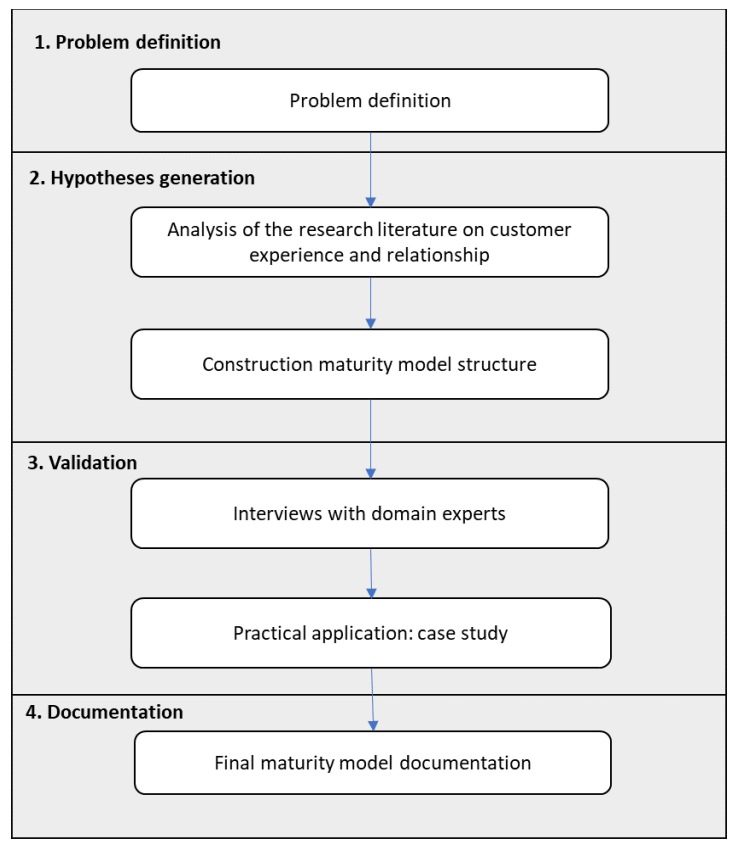

*Figure 1.* Method for designing the maturity model

# Hypothesis generation

## Analysis of the research literature

For defining the maturity model, we carry out a review of the research literature (Webster & Watson, 2002) on DT for improving Customer Experience and Customer relationship by using the Scopus and Google Scholar search engines. The search terms include "digital transformation", digitalization, digitization, digital, "customer experience", and "customer relationship". Then, we undertook several filtering rounds that included reviewing the title, abstract, structure and a complete reading of the papers in order to consider only works addressing improvement of the relationships with customers through digital tools and capabilities. As a result, we found a final set of 28 research works that includes (Adam et al., 2020; Anke et al., 2019; Barann et al., 2020; Chernova et al., 2020; De Keyser et al., 2015; Djurica & Figl, 2017; Gimpel et al., 2016; Hahn, 2019; Hauser et al., 2019; Heuchert et al., 2018; Jamous et al., 2019; Loubochkin & Zasenko, 2020; McLean & Wilson, 2019; Mergel et al., 2019; MIT-Capgemini, 2018; Morakanyane et al., 2017; Mosa et al., 2020; Muhammad et al., 2018; Riedmann-Streitz, 2018; Rode & Stammen-Hegener, 2020; Schaer & Stanoevska-Slabeva, 2019; Stoffer et al., 2018; Vial, 2019; Westerman et al., 2011; Westerman G. et.al., 2014a, 2014b; Yang et al., 2020; Ziaie et al., 2021). In order to extract the concepts necessary for the construction of our framework, we defined review questions which were responded through the revision of the literature as described below.

As said in the introduction, we consider that organizations need to improve the customer experience through all touch points through the whole customer relationship lifecycle process. Consequently, the first review question is: what are the customer relationship lifecycle stages on which maturity needs to be assessed to improve the customer experience? To this question, most of the works respond with several phases, stages and steps that we summarize in four main categories.

**Customer understanding:** it relates identifying the problems, needs and motives that prompts a customer to buy a product or service as well as get customer information, improve understanding of customer preferences, behaviours, tendencies, and enable customization of products or services. Social methods for identifying and gathering such elements include surveys, interviews, observations, focus groups, becoming a user, among others. More technical methods include analysing information traces when the client browses the internet or carry out electronic transactions, which includes going to internal sources of information such as CRM, ERP, digital commerce tools, and soon, as well as to external sources such as browser cookies, external databases that contain financial movements, travel information, hotel stays, etc.

**Customer attracting:** it includes marketing and advertising strategies that help organizations build brand awareness and inform customers about the characteristics and advantages of their products and services. The objective is to attract attention of current or potential customers by using technology tools such as search engine marketing, social networks, e-mail, affiliates schemas and soon. It embraces also traditional technics such as showrooms, television, newspapers and magazines, call center information, etc.

**Selling:** it relates offering products and services that could satisfy customer needs, responding questions of customers regarding the products and services, and finally making effective the transaction in which money is exchanged for a good or service. It can be done through technology means such as e-commerce, websites, social networks, instant messaging, and soon. Traditional means can be also used such as selling through telephone, personal visit to the customer, stores, etc.

**Post-sale service:** it is defined as all activities geared towards maintaining the quality and reliability of the product or the service once it has been delivered with the goal of ensuring customer satisfaction. It may be provided by a retailer, manufacturer, or a third-party customer service or training provider. Depending on the product or service, after-sales activities that include support regarding warranty service, training, or repair and upgrades, can be done through traditional or digital channels.

One of the main impacts of TD initiatives refers to the development of digital capabilities enabled by Information Technology (IT). An IT-enabled capability refers to an organizational capability implemented by using and leveraging IT to differentiate itself from competitors (Sambamurthy & Zmud, 2000). An organizational capacity refers to the ability to acquire or develop knowledge for the coordination of resources, the execution of processes, routines and activities, and the development of the skills that the firm requires. Previous research has found evidence that capabilities enabled by digital technologies lead to better organizational performance (Ayabakan et al., 2017) and is a driver of business value (Benitez et al., 2018). Considering the aforementioned, the second review question is: what are the IT-enabled capabilities that need to be assessed in each stage or category of the customer lifecycle? To this question, we found the following capabilities for each category in the customer lifecycle.

**Customer understanding.** In this category, capabilities allow the organization to obtain customer information, improve understanding of customer preferences, behaviours, tendencies, and enable customization of the products and service offerings. Capabilities for this category are presented as follows:

- *Customer segmentation based on information analysis:* it relates the use of technological tools such as databases and business intelligence applications to classify customers according to information obtained from internal and external sources.
- *Customer sentiments analysis*: this capability refers to the ability of obtaining and analysing customer sentiments and opinions about company's products, services, brands and corporate image by using technological tools.
- *Behaviour and tastes analysis of potential customers*: it describes the ability of the organization to create profiles of current and prospective customers that include their tastes, behaviours and preferences by using technological tools.
- *Management of current customer base with computer systems:* it relates the ability to gather, storage, analyse and manage customers' information through computer tools.
- *Integration of information sources:* it concerns the updating, management and consolidation of multiple customer information sources by using technological tools such as centralized data warehouses or big data solutions.

**Marketing and sales processes.** In this category, capabilities allow the organization to transform the company's revenue channels, improve customer attraction and retention tools as well as optimize related processes by using different technological tools. Capabilities for this category are presented as follows:

- *Use of digital sales channels:* it concerns the implementation and proper use level of digital sales channels, as well as their integration with traditional channels.

- *Use of digital marketing channels: it relates* the implementation and proper use level of digital marketing channels, as well as their integration with traditional channels.
- *Predictive marketing implementation*: it refers the ability of the organization to implement business intelligence and forecasting analysis in marketing processes.
- *Digitalization of operative sales processes:* it concerns the organization's ability to use technological assistance or automation tools to support the sales process to make it more agile and effective for customers.
- *Mobility in the sales process:* it concerns the availability level of sales tools in different mobile devices, which facilitate the relationship with the customer during the sales process.
- *Control on the sales process:* it evaluates the visibility and control level of the sales process by the customer in order to generate a personalized and secure experience.

**Customer service.** In this category, capabilities allow the organization to speed up and be more effective to solve service or post-sale requirements. Capabilities for this category are described as follows:

- *Use of digital channels for customer service:* it concerns the implementation and proper use level of digital service channels.
- *Coherence between the communication channels used with customers:* it is the ability to align and achieve coherence between customer service channels.
- *Implementation of simple and agile technological service tools: it* evaluates the level of simplicity and efficiency of the service tools offered to the customer.
- *High availability of digital service channels:* It refers to the ability of offering high levels of the continuity and availability of service tools.
- *Use of self-service tools for requirements:* it relates to the existence of platforms that allow customers to get access to services by their own and follow up their requirements in a simple and agile way.
- *Service experience feedback channels: it concerns* the availability of feedback channels to gather user experience and feedback in relation to customer service.

To evaluate each capability, maturity models assess related dimensions in the organization in order to estimate the maturity level in which the organization can be allocated. The third review question is thus, which capability dimensions need to be evaluated to determine the maturity level of an organization in its way through the digital transformation of the customer experience? To this question, we found six main dimensions that can help us determine the evolution of an organization in the capabilities previously described: standardization, information, integration, automation, intelligence, and interaction with the context. Those apply in a distinct manner in each capability as it is described in subsection "*The maturity model and reference states definition*".

**Towards a model for digital maturity assessment**

We design the maturity model by using a two-dimensional structure (measurement criteria and maturity levels) which is very common in maturity models designed in the academy and the industry (Santos-Neto & Costa, 2019):

- *Measurement criteria definition:* it is proposed to extract measurement criteria from the digital capabilities important for improving the customer experience described above in the literature review.
- *Maturity levels definition:* it is proposed to define maturity levels that conforms the measurement scale from the Likert scale used in most of the maturity models (Santos-Neto & Costa, 2019). It is composed of 5 maturity levels which will be evaluated for each criterion.

According to the structure adopted, a reference state description in the intersection of each measurement criterion and each maturity level is to be stated, which can help us as a comparative reference during the evaluation process. Reference states are proposed relying on the dimensions extracted from the literature review as well as from the experience of the experts that were interviewed during the development of the model.

The framework elaborated from the literature review and after the first validation activity (interviews with domain experts) is presented in Figure 2. The interviews and respective modifications to the framework from this activity are described as follows.

# Interviews with domain experts

The model in their first version was revised and validated by 4 experts. The validation process was performed by doing personal interviews of one hour with each expert. Previously to each interview, we sent the draft of the first version model to them with a detailed explanation of the design, components, and methodology. During these interviews they freely exposed their observations and doubts about the model and proposed some changes and modifications based on their experience. These proposed modifications were discussed and adjusted during the same interview. Below are presented the profile of each expert and the changes or modifications proposed in the personal interviews:

**Expert 1.**

He is an administrative and commercial manager within an agricultural company dedicated to poultry farming, with more than 20 years of experience in management roles in this industry. He has studies in veterinary medicine and computing systems engineering.

After doing the revision of the proposed model, he considered that the model is pertinent and good structured, but some business capabilities do not apply based on the characteristics of the business industry in which he is working. It is in fact a Business to Business model company in which generally it is not usual to use technology to maintain a relationship with their customers, because their products, considered a raw material, are distributed to a relatively small number of companies, that process it into a final product or resell to the final customer, making no necessary for example using techniques such as sentiment analysis or behaviour and tastes analysis. In this way some of those capabilities of the framework will not be a priority in the short time for him.

As a result, he proposed that the framework could include assigning different weights for each criterion capabilities or capability in order to make it more flexible in the application to different industries. This recommendation was implemented as a change in the application methodology of the framework, including in the final version, the possibility to weight each business practice into each criterion, to reduce the importance of the business capabilities that are not relevant or augment those that are important according to the nature of the business that is being evaluated.

**Expert 2.**

He is a technical and presales manager in a multinational technology company, with more than 10 years of experience working in technical, commercial and management roles withing the technology industry.

He found clear the evaluation framework and considered it aligned to the sales process and customer relationship capabilities that technology companies are used to implement. He argued that in the business practice "Mobility in the sales process", the most advanced maturity level should involve the complete adoption by the sales team of the company of mobile technology tools and apps for enterprise systems such as CRM, ERP and product catalogs. He supports this suggestion by considering that in his experience the correct adoption of such technologies in the day-to-day work should be considered as something demonstrating the maturity of a company in this aspect. This recommendation was included in the reference states of the highest levels of maturity in the framework, where previously was not considered the correct and complete adoption of the technological tools in the day-to-day work.

**Expert 3.**

He works as a business development manager in a multinational technology company. He has more than 16 years of experience in operations management, customer service management, project management and business development, and holds a bachelor's degree in electronics engineering and has a master's degree in project management.

He found that artificial intelligence technologies are not mentioned in the reference states of highest maturity level of the business capabilities belonging to the customer understanding criterion (e.g., Customer segmentation, sentiment analysis and behaviour and taste analysis). For this reason, he suggested to include the implementation such technologies in the most advanced maturity levels, keeping in mind that those technologies are useful to automate and agile the customer analysis process. Also, he suggested and helped us to define in detail the reference states for advanced and partial implementation of each business capabilities of the customer service criterion. From this recommendation, it was included that in the reference states for levels 2, 3, and 4 of the business capabilities of customer service criterion, the use of some traditional service channels as e-mail and telephone, that were not in the draft version, and also it was considered the knowledge base concerning solved cases repository for an intermediate level of the self-service tools capability.

Finally, he suggested to include into the methodology, a presentation of a formal report of the maturity assessment to the board of directors, as a making decisions source for each company that applies this proposed maturity model. This recommendation was implemented in the step 5 of the proposed application methodology of the assessment model.

**Expert 4.**

She is a regional sales manager in a technology company, with more than 20 years of experience in different regional and multinational companies in roles of general managing director, sales director, and operations director. She has studies in computer engineering and a post-graduate degree in senior management.

She suggested to expand the former criterion called "Sales Process" to call it "Marketing and Sales process". She argued that this criterion includes in reality capabilities for both processes and, as a consequence, it must be clear that marketing is also addressed there. In addition, she suggested to add in this criterion the capability "Use of electronic marketing channels" that was not included in the first model version, and it would make sense, to do a detailed evaluation of the alignment with the customers in this aspect that is essential in multichannel business of today.

She finally suggested to change the name of the business practice initially called "Visibility on the sales process", to "Control on the sales process" within the "Marketing and sales process" criterion, as in her opinion it defines better the objective of this business practice, which is to evaluates how customers can follow up and control the sales process in their benefit, so that the word "control" is more accurate than "visibility" and is more understandable to companies in different industries.

## Refinement of the model

### Model structure

As previously said, we design the maturity model by using a two-dimensional structure (measurement criteria and maturity levels) which is very common in maturity models designed in the academy and the industry (Santos-Neto & Costa, 2019):

**Digital capabilities definition.**

The definition of measurement criteria and business capabilities (see Figure. 2) result from the literature review, and then refining by the recommended modifications from the expert's interviews (see section "*Analysis of the research literature*").

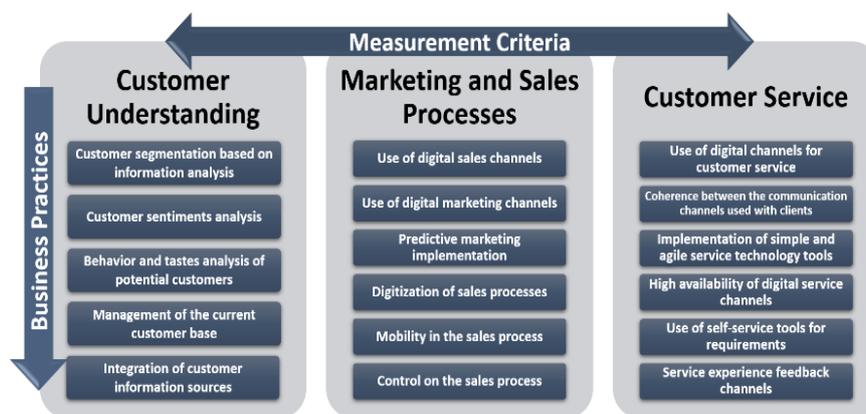

Figure 2. Customer alignment maturity measurement model

**Maturity levels definition**

It is proposed to assess measurement criteria under the following 5 maturity levels:

1. *Initial / Process Ad Hoc: the p*ractice is not implemented or not conceived to help organizations align with customers.
2. *Committed process:* the organization has plans to implement and/or improve the practice.
3. *Focused and stabilized process: t*he practice is established but external alignment with customers is still lacking.
4. *Improved / Managed Process: w*ithin the practice digital and disruptive technologies are conceived as a value element towards the alignment with customers.
5. *Optimized Process:* the practice has been entirety implemented, is flexible in the face of external market changes and helps the organization gain competitive advantages.

# The maturity model and reference states definition

Specific reference states are established for each practice within each maturity level, which will serve as a guide for the maturity assessment. The complete maturity model is presented in Table 1.

Table 1: Maturity model for customer alignment

| C | Practice | Level 1 | Level 2 | Level 3 | Level 4 | Level 5 |
|---|---|---|---|---|---|---|
| Customer understanding | Customer segmentation based on information analysis. | There is no segmentation of the customer base. | Customers are inaccurately segmented from incomplete information, however, there are plans to improve the information sources and analysis tools | Customers are segmented based on local data analysis. Analysis tools are obsolete or has limited functionalities and sources. | Customers are segmented based on local data analysis. CRM tools fully implemented managed by the IT area. | Customers are segmented based on analysis of local and external data. CRM and business intelligence tools are fully adopted and managed by the IT and business areas. Also exist some artificial intelligence intervention. |
| Customer understanding | Customer sentiments analysis. | There are no tools or data sources for customer sentiment analysis. | Manual monitoring and analysis are conducted from a single source of information. | Monitoring and semi-automatic analysis of several sources with isolated tools for each source. | Semi-automatic monitoring and analysis is performed consolidating data from multiple sources. | Automatic monitoring and analysis are performed consolidating data from multiple sources, using filters and artificial intelligence. |
| Customer understanding | Behaviour and tastes analysis of potential customers. | There are no tools or data sources for analysis of behaviour and preferences. | Manual monitoring and analysis are done with local and limited data sources. The intuition is the major guide. | Semi-automatic monitoring and analysis is done with some data capture tools in web portals and social networks. Free limited online tools are used. | Automatic monitoring and analysis are done with centralized tools and multiple web sources of information. | Automatic monitoring and analysis are done with centralized and intelligent tools and multiple webs and IoT sources. Predictive modeling is possible. |
| Customer understanding | Management of current customer base with computer systems. | There is no database of current customers | Customers are managed with an outdated database. Uncontrolled and messy spreadsheets are used. | Customers are managed in a local database with occasional updates. Is used and administered a simple database engine. | Customers are managed with a CRM system with little adoption and inadequate use. | Customers are managed with an updated CRM system in real time through mobile and local applications. |
| Customer understanding | Integration of information sources of current customers and prospects. | There is no integration strategy for information sources. | There are isolated sources of information with plans for future integration. | The information sources are partially integrated and are manually updated with low frequency. | There is a central data bank that is frequently updated with information from a limited number of sources. An ETL tool and strategy are implemented. | There is a central data bank that is updated in real time with customer information through multiple channels. An Enterprise Data Warehouse and/or Big Data environment is fully implemented. |
| Marketing and sales | Use of electronic sales channels. | There are no electronic sales channels. | There is a catalog of products within the web portal | The web portal implements an online sales platform with some products and a complex process. | The web portal implements an online sales platform of its complete portfolio and with a simple process. Managed by the IT and sales area. | There are multiple electronic sales channels through social networks and the web portal, managed by the IT and sales area. The sales channels are integrated and coherent. |
| | Use of electronic marketing channels | There are no electronic marketing channels | There are plans to implement electronic marketing media, emails, and web pages. | The email is used to send marketing material to customers. There is no monitoring of its effectiveness. | Email and social networks are used as marketing channels and there is manual monitoring of effectiveness. | Consolidation tools of electronic marketing channels are used, and full monitoring of the effectiveness of the campaigns is made. |
| | Predictive marketing implementation. | There is no predictive marketing | Marketing based on limited customer information | Marketing based on full local customer information based on manual descriptive analysis. | Marketing based on trends of local and external customer information based on descriptive analysis tools. | Marketing based on predictive models in artificial intelligence tools and with local and external customer information. |
| | Digitalization of operative sales processes towards customers. | Completely manual sales processes | Manual sales process with digitization plan in progress | Partially digitized sales process with electronic payment | Digitized sales process with paper support documents | Sales process completely digital and connected to the company's information systems. |
| | Mobility in the sale process. | There are no sales tools accessible through mobile devices | There are electronic sales tools with projected implementation of mobile access | There are sales tools with the possibility of mobile access, but without implementation of their use | Sales tools have mobile access implemented. | Sales tools have mobile access implemented and real-time connection to information systems. Completely adopted by the sales team. |
| | Control on the sales process. | The sales process is not visible to the customer | The sales process is partially visible to the customer. Some manual notifications during the sale. | The sales process is visible to the customer in its entirety without granting control to the client. (tracking) | The customer has visibility and partial control of the sales process. (Ex. Times and delivery places can be managed) | The sales process is visible and customizable by the customer. (Eg. Shopping channels, offers, times, deliveries) |
| Customer service | Use of digital channels for customer service. | There are no digital channels for customer service | There is an implementation plan of the customer service channel by e-mail. | Customer service is provided through the contact section of the web portal and through telephone line. | Customer service is provided through different digital channels such as social networks, e-mail and web portal | There is a centralized management platform for customer service through social networks, e-mail and a web portal with information available to customers. |
| | Coherence between the communication channels used with customers. | The communication channels with customers do not share information between them | There is a plan to integrate information from traditional communication channels such as email, telephone, and web portal | Traditional communication channels share information with low frequency | The traditional communication channels are integrated and coherent by the management of the IT area. The new channels have partial integration. | Traditional communication channels are integrated and coherent with new channels such as social networks. The IT area has visibility and total technical control. |
| | Implementation of simple and agile technological service tools. | There are no technological tools for customer service | There are plans to implement traditional technological tools for customer service. | There are obsolete technological tools for customer service with high levels of complexity | There are agile customer service technology tools as social networks, remote service, etc. with multi-channel integration | There are advanced customer service tools with some level of autonomy as chatbots. |
| | High availability of digital service channels. | Customer service channels do not have high availability strategies | There is a project to implement high availability strategies in service channels | The main service channel has high availability, but is not implemented for all customer service channels | Traditional customer service channels have high availability. No new service channels are contemplated. | It has high availability for traditional service channels and is integrated with channels in the cloud as social networks. |
| | Use of self-service tools for requirements. | No self-service requirements tools are implemented | The service request tools have a projected implementation of self-service modules. | Service request tools allow self-service partially for clients. There are a solved cases repository accessible by clients. | There is a complete self-service platform for customer requirements with occasional assistance by the service team. The platform automatically recommends past solutions. | There is a complete self-service platform for customer requirements assisted by intelligent systems that uses heuristics. Minimum human assistance. |
| **Service experience feedback channels.** | Service experience feedback channels are not implemented. | There is a basic survey on unconsolidated paper. There is a plan to implement digital channels. | A service comment box is available through a web portal or email. | The service tools implement digital feedback channels associated with the requirements. (Online forms) | There is multi-channelity for the feedback of the service experience. The channels are integrated and relate to the requirements of each client. Social networks are listened. |

**Assessment model application methodology**

For the application of the model, it is necessary to define an evaluation team made up of business and IT managers and executives, who are in charge of answering and discussing the maturity level that fit the best to the current state of the company in each of the criteria and business practice. This group define the gaps found and the possible actions to be taken from the results. It is recommended to include an external consultant to guide the assessment process.

The specific steps for the application of the measurement model considers de method proposed in (Luftman, 2003), that includes the following 5 steps:

1. *Conform the evaluation team:* to create a team of IT and business executives for the evaluation. The number of executives may vary depending on the size of the company and whether a business unit or the entire organization is evaluated.
2. *Gather relevant information:* The defined team must evaluate each of the business capabilities. This can be done in three ways:
    a. As joint work of the team through different meetings.
    b. Responding to a survey by each of the members and then organizing a meeting to discuss and consolidate results.
    c. Combining the first two methods, it means part of the evaluation team could work in a presential meetings and the others could do it by answering online surveys.
3. *Decide individual scores:* The team must reach an agreement to assign a score to each of the capabilities, highlighting the gaps found and the possible steps to be followed to solve them. The reference score for each of the criteria will be the average of the scores of the capabilities that it groups.

4. *Decide the overall maturity score:* The team must achieve a consensus on the final score assigned to the current state of the company. The average scores of each criterion will guide this consensus, but the team can adjust the total score if it considers that certain capabilities have more or less weight within the company than others, according to their industry context.
5. *Present an executive report:* After consolidating the partial scores and the final score, it is recommended to prepare an executive summary of the evaluation for the executive board, which includes maturity levels, gaps and possible improvement strategies. It is proposed to use a general evaluation consolidation format as a reference.

## Validation: practical application

**Case study context**

The maturity model and methodology exposed in the previous section was applied in three companies from different industries, sizes, and business models that operate in the Latin America region. The first subsection presents the profile of the companies (industry, economic activity, approximate quantity of customers base, quantity of employees and the profile of the evaluation team). The second subsection describes the application of the model in each company and the analysis of the results obtained.

**Profiles of participating companies**

Below are presented the profiles of the participating companies in the case study: (see Table 2.)

Table 2. Profiles of participating companies.

| **Company 1:** | |
|---|---|
| Industry: | Hospitality |
| Quantity range of employees: | 50 to 200 |
| Economic activity description: | Hotel and spa services, focused on luxury experiences for long stays to executive customers in Bogota, Colombia. |
| Approximately quantity of clients: | 3.100 |
| Evaluation team profile: | Conformed by 4 executive managers. |
| **Company 2:** | |
| Industry: | Information technology services |
| Quantity range of employees: | 50 to 200 |
| Economic activity description: | It is a business unit of a multinational company of technology and services, in charge to perform the presales and sales of enterprise technology solutions in the Latin America region. |
| Approximately quantity of clients: | 20.000 |
| Evaluation team profile: | Conformed by 6 managers of the business. |
| **Company 3:** | |
| Industry: | Marketing and consultancy services |
| Quantity range of employees: | 1 to 50 |
| Economic activity description: | Company of market research and consultancy in marketing strategy. |
| Approximately quantity of clients: | 100 |
| Evaluation team profile: | Conformed by 5 executive managers of the company. |

We can observe that presented companies have so different profiles in industry and size, what help us to have a more heterogeneous experience in the maturity model application.

**Application of the measurement model**

According to the methodology of application of the measurement model, there are two stages to obtain results: the first is the average result of the individual evaluations, made by the members of the evaluation team. For this stage we sent online forms with the criteria, business capabilities to be evaluated in the 5 maturity levels with the help of the reference states. The second concerns results consolidated under consensus of the entire evaluation team with the identification of gaps and possible future strategies. In this article we will only present the consolidated results.

Table 3 presents the reference codes for each criterion and business capabilities which will be used in the presentation of the evaluation results.

Table 3. Criterion and business capabilities reference codes.

| Criterion | Practice | Practice code |
|---|---|---|
| A. Customer understanding | Customer segmentation based on information analysis. | A.1 |
| | Customer sentiments analysis. | A.2 |
| | Behaviour and tastes analysis of potential customers. | A.3 |
| | Management of current customer base with computer systems. | A.4 |
| | Integration of information sources of current customers and prospects. | A.5 |
| B. Sales and Marketing processes | Use of digital sales channels. | B.1 |
| | Use of digital marketing channels. | B.2 |
| | Predictive marketing implementation. | B.3 |
| | Digitalization of operative sales processes towards customers. | B.4 |
| | Mobility in the sale process. | B.5 |
| | Control on the sales process. | B.6 |
| C. Customer service | Use of digital channels for customer service. | C.1 |
| | Coherence between the communication channels used with customers. | C.2 |
| | Implementation of simple and agile technological service tools. | C.3 |
| | High availability of digital service channels. | C.4 |
| | Use of self-service tools of requirements. | C.5 |
| | Service experience feedback channels | C.6 |

Below, the final consolidated results, corresponding final score, main identified gaps, and analysis for each company.

**Company 1 results and analysis:**

Table 4 presents average score, weight, average score by criterion and final score for the first company. The evaluation team decided to maintain the obtained scores after the average of the individual evaluations made by themselves, as well as to conserve the same weight for all the capabilities into the evaluation.

Table 4. Results by consensus of evaluating team of Company 1

| Criterion | Practice | Average level | Weight % | Average by criterion |
|---|---|---|---|---|
| Customer understanding | A.1 | 1,3 | 20% | 1,9 |
| | A.2 | 3,0 | 20% | |
| | A.3 | 1,8 | 20% | |
| | A.4 | 2,3 | 20% | |
| | A.5 | 1,3 | 20% | |
| Sales and Marketing processes | B.1 | 2,5 | 16,67% | 2,1 |
| | B.2 | 2,5 | 16,67% | |
| | B.3 | 1,3 | 16,67% | |
| | B.4 | 3,3 | 16,67% | |
| | B.5 | 1,5 | 16,67% | |
| | B.6 | 1,5 | 16,67% | |
| Customer service | C.1 | 1,5 | 16,67% | 1,4 |
| | C.2 | 1,0 | 16,67% | |
| | C.3 | 1,8 | 16,67% | |
| | C.4 | 1,0 | 16,67% | |
| | C.5 | 1,0 | 16,67% | |
| | C.6 | 2,0 | 16,67% | |
| **FINAL SCORE** | | | **1,8** | |

*Main gaps identified.*

- In customer understanding, it is identified the lack of a database of current customers with complete and relevant information and a collecting information system to get or update customer data. Currently they do a manual management of customers information which is inefficient.
- In the sales and marketing process, it is identified that current tools are limited and not connected with a customer database. In addition, there is no digital sales and marketing channels, as the only contact points for this activity are personal meetings and telephone conversations.
- The customer service and post-sale communication is performed by a manual survey. Additionally, the company maintain social network pages, but they are not used adequately as customer service or communication channels.

*Results analysis:*

This company is relatively young with 15 years in the market, but it operates under a traditional business model where the technology continues to be an operative enabler with limited implementation. According with the results, it is confirmed the diagnostic of lack of a technology strategy, specifically, in the process of relationship with customers lifecycle (pre-sales, sales, and post-sales).

This company obtained a maturity level of 1.8 (it can be considered as near to level 2). They are aware of their low maturity level in their customer alignment but have plans to improve their marketing and sales processes implementing technological tools. As improvement strategy, it is proposed to make structured plans to gain capabilities in the customer understanding area through the implementation of technological tools such as databases and traditional CRM systems.

**Company 2 results and analysis:**

Table 5 presents average score, weight, average score by criterion and final score for the second company. The team members maintained the average scores obtained from the individual evaluations. Regarding the weight of the evaluated capabilities, relevance to customer sentiment analysis was removed because within a B2B business model, it is not of vital importance. The remaining capabilities maintained an equal weight within the evaluation.

Table 5. Results by consensus of evaluating team of Company 2

| Criterion | Practice | Average level | Weight % | Average by criterion |
|---|---|---|---|---|
| Customer understanding | A.1 | 4,2 | 25% | 3,54 |
| | A.2 | 2,2 | 0% | |
| | A.3 | 2,8 | 25% | |
| | A.4 | 3,8 | 25% | |
| | A.5 | 3,3 | 25% | |
| Sales and Marketing processes | B.1 | 3,3 | 16,67% | 3,3 |
| | B.2 | 4,0 | 16,67% | |
| | B.3 | 2,7 | 16,67% | |
| | B.4 | 3,5 | 16,67% | |
| | B.5 | 3,7 | 16,67% | |
| | B.6 | 2,8 | 16,67% | |
| Customer service | C.1 | 4,0 | 16,67% | 3,3 |
| | C.2 | 2,8 | 16,67% | |
| | C.3 | 3,5 | 16,67% | |
| | C.4 | 3,2 | 16,67% | |
| | C.5 | 4,0 | 16,67% | |
| | C.6 | 2,5 | 16,67% | |
| FINAL SCORE: | | 3,4 | | |

*Main gaps identified.*

- In the customer understanding criterion, although the company has advanced CRM tools, there is no full and adequate use of all its functionalities for customer relationship. Problems of information quality and source integration are also identified.

- In sales and marketing, despite of its Business to Business (B2B) model, it is desirable to implement more agile technological tools to allow final customers to enter and track purchase orders. Mobile tools for sales exist, but a full and adequate use is not achieved due to the lack of management from the managerial level. There are plans to implement artificial intelligence within the marketing processes, but it is not yet available for use within the evaluated business unit.

- In customer service and post-sale communication, technological tools lack self-service options and there are problems of availability of resources to respond quickly to support post-sales customer requests. In addition, there is not a user experience feedback channel. The implementation of artificial intelligence was identified as an improvement option to assist customers in the post-sales support processes.

*Results analysis:*

As a business unit of a multinational technology company with a presence and a long history throughout Latin America, the operating characteristics of this unit are quite mature and with a very good technology adoption level established within its business strategy. Even so, there are significant gaps in the alignment with customers as observed in the results obtained, for this reason a maturity level of 3.4 was assigned to this company, what is somewhat higher than 3. The unit has thus capabilities for using technological tools with a higher level in customer understanding and, in the sales process, and this positions the company, in general, in an advanced range of maturity.

Among the proposals for the customer experience improvement, one of the most important is the evaluation and projection of using new and disruptive technology such as artificial intelligence, as well as the implementation of information systems integration processes. This would help reach a maturity level 4.

**Company 3 results and analysis:**

Table 6 presents average score, weight, average score by criterion and final score for the third company. This company made some adjustments to the individual evaluations at the assessment board meeting, where they considered that their initial evaluation had been very low. The weight of capabilities A.2, B.5, B.6, C.3 was also reduced to zero because, according to their evaluation of the business model context, for their case they were not important for customer relationship.

Table 6. Results by consensus of evaluating team of Company 3

| Criterion | Practice | Average level | Weighting in% | Average by criterion |
|---|---|---|---|---|
| Customer understanding | A.1 | 4,0 | 30% | 2,8 |
| | A.2 | 3,0 | 0% | |
| | A.3 | 3,0 | 20% | |
| | A.4 | 2,0 | 25% | |
| | A.5 | 2,0 | 25% | |
| Sales and Marketing processes | B.1 | 2,0 | 10% | 2,7 |
| | B.2 | 3,8 | 40% | |
| | B.3 | 1,0 | 20% | |
| | B.4 | 2,6 | 30% | |
| | B.5 | 1,6 | 0% | |
| | B.6 | 2,0 | 0% | |
| Customer service | C.1 | 3,5 | 20% | 3,5 |
| | C.2 | 4,0 | 20% | |
| | C.3 | 2,6 | 0% | |
| | C.4 | 4,0 | 20% | |
| | C.5 | 2,0 | 20% | |
| | C.6 | 4,0 | 20% | |
| **GENERAL LEVEL:** | | | **3,0** | |

*Main gaps identified.*

- In customer understanding, there is a lack of tools for customer behaviour and taste analysis, so that it is still a reactive sales process.
- In the marketing and sales, it is identified a lack of digital sales channels and product portfolio management based on the past performance products and services. Also, there is a lack of assistance tools in the sales process to guide the customers. Due to the lack of information of customer behaviours and tastes, there is no predictive marketing strategy neither.
- In customer service, there are different communication channels such as chat, e-mail, and videoconference systems, but it does not exist an omni-channel management strategy for them. Also, it is identified the lack of an online information system allowing the customer to follow the projects that are in execution, as well as an online customer experience feedback channel.

*Results analysis:*

Due to the nature of its business, which involves a relatively low number of customers, it is a company that has a high knowledge of its customers. Their evaluation in post-sales customer service is high even when the participants identified important gaps in the use of technological tools, which generates a kind of doubt in the evaluation of this criterion.

The general evaluation of maturity is 3.0, positioning the company at the level of focused and stabilized processes within the evaluation scale. From this evaluation, it can be considered that even when it is a company with a business model focused on enterprise customers, the implementation of electronic sales tools for standardized products can improve its operation. Additionally, the implementation of flexible customer service channels with high availability, as well as a user experience feedback channel, is recommended as a critical point of improvement.

**Results comparison**

To have a reference comparison of the results for these three companies, here is presented a graph which shows the gaps between them: (See Figure. 3)

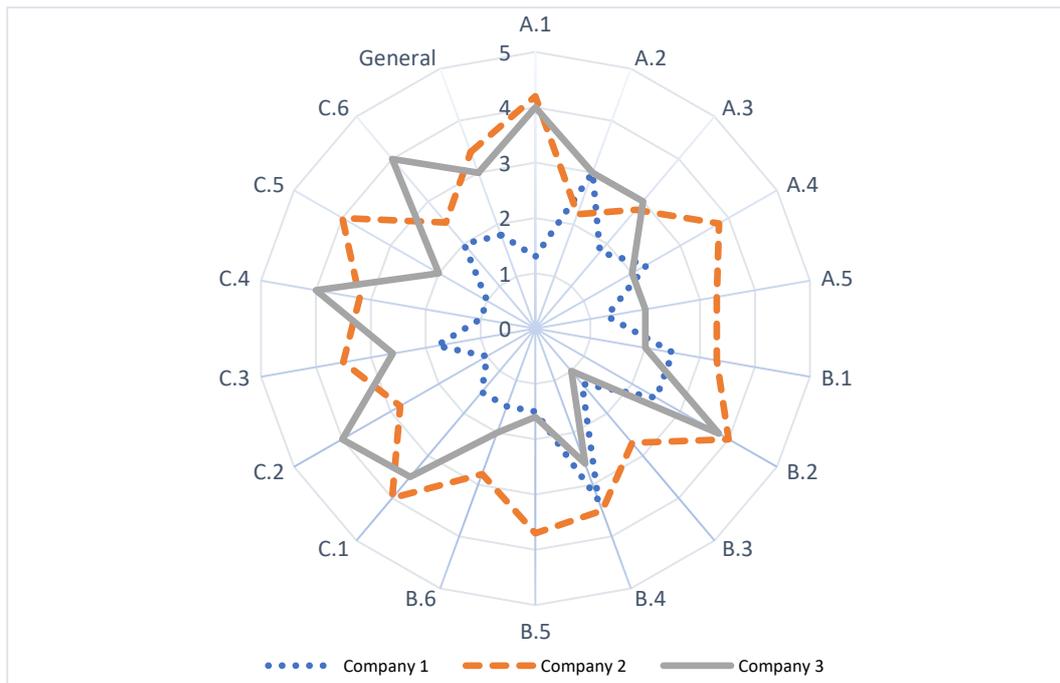

Figure 3. Graphical comparison of the results from the three companies

## Discussion

Comparing our work with existing maturity and capability models in the digital transformation domain, the first conclusion is that there are no specific proposals to assess digital capabilities supporting the whole customer lifecycle. There is only one work specialised in assessing digital marketing performance (Chaffey, 2010) that covers aspects such as online customer acquisition and conversion, customer experience, cross channel integration, and brand development among others. However, this work has the following shortcomings which we try to cover with our model. On the one hand, the approach is focused on digital capabilities for marketing and presales and does not cover the whole customer relationship lifecycle, which prevents organisation to assess crucial capabilities to offer a complete experience to customers. On the other hand, it does not consider last advances in IT such as social technologies, mobile devices and artificial intelligence and is only focused on the corporate website, which is not consistent with current customer's behaviour and use of technology.

Regarding more general frameworks in domain, on one side, some of them (Amaral & Peças, 2021; Barry et al., 2022; Gökalp & Martinez, 2021; Santos & Martinho, 2020) are focused on the assessment of aspects such as strategy, organisation, structure, culture, among others, that are managed by companies to contribute to the Industry 4.0 and DT maturity from an structural and internal viewpoint. Consequently, such works do not define criteria or parameters to assess capabilities helping organisation improve its relationship with external customers. On the other side, frameworks that consider external or contextual dimensions such as use of customer data, distribution channels and connectivity, customer journey analysis, or customer's digital media competence have the following shortcomings: first, some of them (Gollhardt et al., 2020; Schumacher et al., 2016) do not present reference states to guide the maturity level assessment for each evaluation criteria which is one of the main contributions of our work, and are not completely described in terms of their criteria or assessment scales which does not allow researchers and practitioners to use them in real business scenarios, second, other (Castelo-Branco et al., 2022; Gollhardt et al., 2020; Soares et al., 2021) have not been applied through case studies so that their weakness and strength cannot be evaluated nor verified.

## Conclusions and future work

The current business environment is constantly changing and currently the change frequency is higher, for this reason companies are aware of the need of improving the customer experience by making many efforts and investing resources in the adoption of new technologies within their business models. A qualitative model was developed to assess the digital capabilities supporting the transformation of the customer experience. From the validation of this model through experts' interviews and three case studies, it was concluded that:

(i) Although the measurement model has reference states at each level for each capability, it cannot cover all the business models and industries in a generic way due to the high complexity that each context implies.

(ii) The qualitative nature of the maturity measurement model allows a high level of subjectivity in the evaluation, which needs to be mitigated through a group discussion meeting of the evaluation team.

(iii) Although companies' maturity levels are low and medium due to the lack of formality in customer experience tools and processes, thanks to the assessment the companies learnt to be aware of the importance of this aspect.

(iv) It can be concluded that to make the evaluation process more precise, it is necessary to include an external business consultant that helps generate a more neutral environment in the assessment process, further reducing the subjectivity of the application of the maturity measurement model.

(v) Despite been a model with some subjectivity level, its application allows companies to do a structured and guided evaluation of their digital capabilities for customer experience and a base roadmap to the next steps for a higher maturity level. This should be so difficult to achieve only with focal groups without a structured model.

The following important topics were identified for future work. It is necessary to validate and adjust the maturity model by applying it to more different industries and obtain feedback to adjust the criteria and capabilities evaluated, if possible, by following a contingent approach dependent on each industry. It is necessary to propose strategies in coherence to the model in order to allow companies to go from a current maturity state to the next one.